\documentclass[preprint, aps,amsmath,amssymb,floats,12pt]{revtex4}
\usepackage{amsmath}
\usepackage{pdfpages}
\usepackage{graphicx}
\usepackage{dcolumn}
\usepackage{bm}
\newcommand{\ket}[1]{\left| #1 \right>} 
\newcommand{\bra}[1]{\left< #1 \right|} 
\newcommand{\braket}[2]{\left< #1 \vphantom{#2} \right|
 \left. #2 \vphantom{#1} \right>} 
\begin{document}

\title{Waxman's Algorithm for non-Hermitian Hamiltonian Operators}
\author{ S. R. Chamberlain}\email[ ]{cham8134@fredonia.edu}\author{ J. G.
Tucker}\email[ ]{tuck5702@fredonia.edu}\author{J. M. Conroy}\email[ 
]{justin.conroy@fredonia.edu} \author{H. G.
Miller}\email[ ]{hgmiller@localnet.com}
\affiliation{Department of Physics, The State University of New York at 
Fredonia, 
Fredonia, NY, 14063}

\begin{abstract}
 An algorithm for finding 
the bound-state eigenvalues and eigenfunctions of a Hermitian Hamiltonian 
operator  using 
Green's method, developed by Waxman\cite{W98},
has been extended to include non-Hermitian Hamiltonian operators.

\end{abstract}
\maketitle 

Non-Hermitian Hamiltonian operators have played an important role in many 
fields of physics. In nuclear physics, optical model calculations as well as 
Gamow  shell model calculations have long been of interest in describing states 
in the continuum. Recently, a Gamow shell model description of weakly bound 
systems in neutron-rich nuclei involving configuration mixing in a single 
particle Berggren basis\cite{B68} has been given\cite{MNPB02}. The Berggren 
basis contains bound  single particle states as well as narrow resonances and 
non-resonant continuum. The Hamiltonian to be diagonalized in this basis  is 
non-Hermitian.

The Waxman Algorithm\cite{W98} is an iterative method based on Green's method 
that 
allows one to determine  eigenstates of a Hamiltonian operator  
without matrix diagonalization.  Note Green's method 
may be applied to Hermitian as well as non-Hermitian operators. 
In the Waxman Algorithm approach,  the coupling constant of 
the potential $\lambda$ is determined numerically as a function of the 
eigenvalue, $\varepsilon$. $\varepsilon$ is then varied until one obtains the 
value 
of $\lambda$ used in the Hamiltonain operator. For non-Hermitian Hamitonian 
operators $\varepsilon$ may be a complex number and an iterative algorithm is 
required to determine the complex eigenvalue corresponding to the 
real value of $\lambda$ used in the Hamiltonian operator.               

Consider the following eigenvalue problem
\begin{equation}
(\hat{T}-\lambda \hat{V})\ket{u}=\varepsilon\ket{u},
\label{geg}
\end{equation}
where $\hat{T}$ is the kinetic energy operator, $\lambda$ is the real coupling 
constant, 
$\varepsilon$ is the energy eigenvalue, and $\hat{V}$ is the potential energy 
operator. For 
non-Hermitian potentials the  energy eigenvalues will in general be complex.  For bound states the solution of Eq. (\ref{geg}) via Green's Method yields
\begin{equation}
\ket{u}=\lambda \hat{G}_{\varepsilon}\hat{V}\ket{u}
\label{green}
\end{equation}

where the Green's operator, $\hat{G}_{\varepsilon}$, is defined as
\begin{equation}
\hat{G}_{\varepsilon}=(\hat{T}-\varepsilon)^{-1}
\end{equation}

and the vector $\ket{u}$ is normalized with a reference vector $\bra{r}$ such 
that
\begin{equation}
\braket{r}{u}=1
\end{equation}

With this, $\lambda$ can be written as
\begin{equation}
\lambda=\bra{r}\hat{G}_{\varepsilon}\hat{V}\ket{u}^{-1}.
\label{lambda} 
\end{equation}
Eq. (\ref{lambda}) can now be substituted into Eq. (\ref{green})
\begin{equation}
\ket{u}= 
\frac{\hat{G}_{\varepsilon}\hat{V}\ket{u}}{\bra{r}\hat{G}\hat{V}\ket{u}}.
\label{u}
\end{equation}
For a chosen value of $\varepsilon$, Eq. (\ref{u}) can be iterated
\begin{equation}
\ket{n+1}= 
\frac{\hat{G}_{\varepsilon}\hat{V}\ket{n}}{\bra{r}\hat{G}_{\varepsilon}\hat{V}
\ket{n}}
\end{equation}
until a convergent solution is obtained, at which point $\lambda$ can be  
determined from Eq. (\ref{lambda}). 

If $\varepsilon$ is chosen to be 
complex, $\lambda$ determined from Eq. (\ref{lambda}) will not necessarily be real. Using polar coordinates where $\lambda=|\lambda|e^{i\phi(\lambda)}$ ,$ |\lambda|=\sqrt{Re[\lambda]^{2}+Im[\lambda]^{2}}$, and $ \phi(\lambda)=\arctan{\frac{Im[\lambda]}{Re[\lambda]}}$,
Waxman's proof 
of convergence implies that there will be convergence
of $|\lambda|$ to the magnitude of the chosen real value of $\lambda$, 
$\lambda_{ex}$, but convergence is not 
guaranteed to be a real solution. To converge to the real solution 
$\lambda_{ex}$, i.e. 
where $\phi(\lambda)$=0, the following method was developed.

For a matrix whose ground state is complex, an arbitrary value of 
$\varepsilon=|\varepsilon|e^{i\phi(\varepsilon)}$ and a
corresponding arbitrary eigenvector are chosen. 
$|\varepsilon|$ is then varied incrementally until 
$|\lambda|$ is within a small range close to the magnitude of the chosen real 
$\lambda_{ex}$.   FIGS. 1 and 2 show $|\lambda|$ vs. $|\varepsilon|$ for a sample 20 x 20 non-Hermitian Hamiltonian Matrix whose lowest lying eigenvalue is complex.
 One can see that $|\lambda|$ and $|\varepsilon|$ are related linearly and varying $\phi(\varepsilon)$ causes a veritcal shift in $|\lambda|$.
\vspace{4mm}

\begin{figure}[htbp]
\centering
 \includegraphics*[scale=1, angle=0]{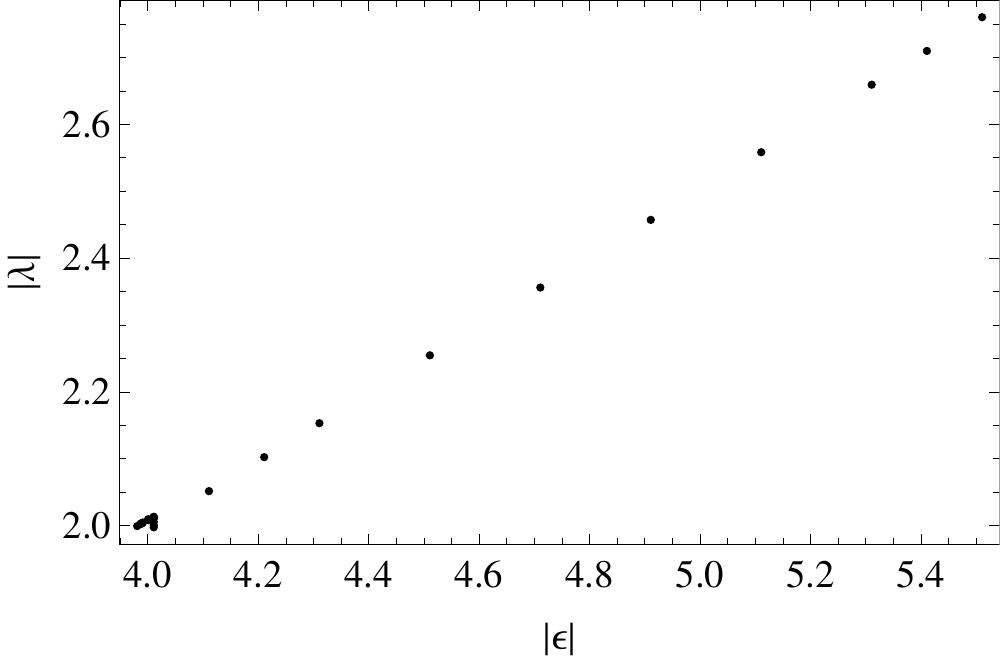}
   \vspace{-5mm}\caption{$|\lambda|$ vs. $|\varepsilon|$ for a 20x20 Hamiltonian matrix with complex ground state eigenvalue.  Convergence occurs at $|\varepsilon|=3.982$ and the chosen real $\lambda_{ex}=2$. See Fig. 2 for iteration near convergence.}
\end{figure}

 \begin{figure}[h!tbp]
  \centering
  \includegraphics*[scale=1]{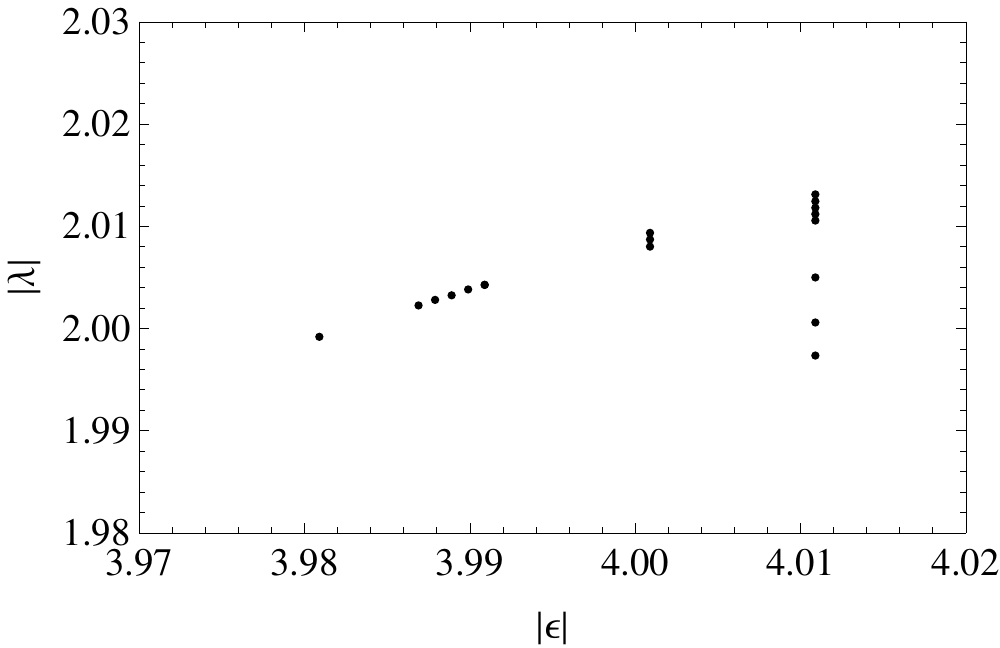}
   \caption{Zoomed-in version of FIG. 1.
Convergence occurs at a value of 3.982 for $|\varepsilon|$.}
\end{figure}

 At this point, $\phi(\varepsilon)$ can then be varied until either $|\lambda|$ 
is 
no longer 
within range of the magnitude of $\lambda_{ex}$, in which case the previous 
step is repeated, or until $\phi(\lambda)=0$, in which case the equation has 
been solved. FIG. 3 shows $\phi(\varepsilon)$ vs.$\phi(\lambda)$, which are also 
linearly related.  When $\phi(\varepsilon)$ is varied, this 
shifts $|\lambda|$ vertically, as seen in FIG. 2.

\vspace{5mm}
\begin{figure}[htbp]
  \centering
   \includegraphics*[scale=1]{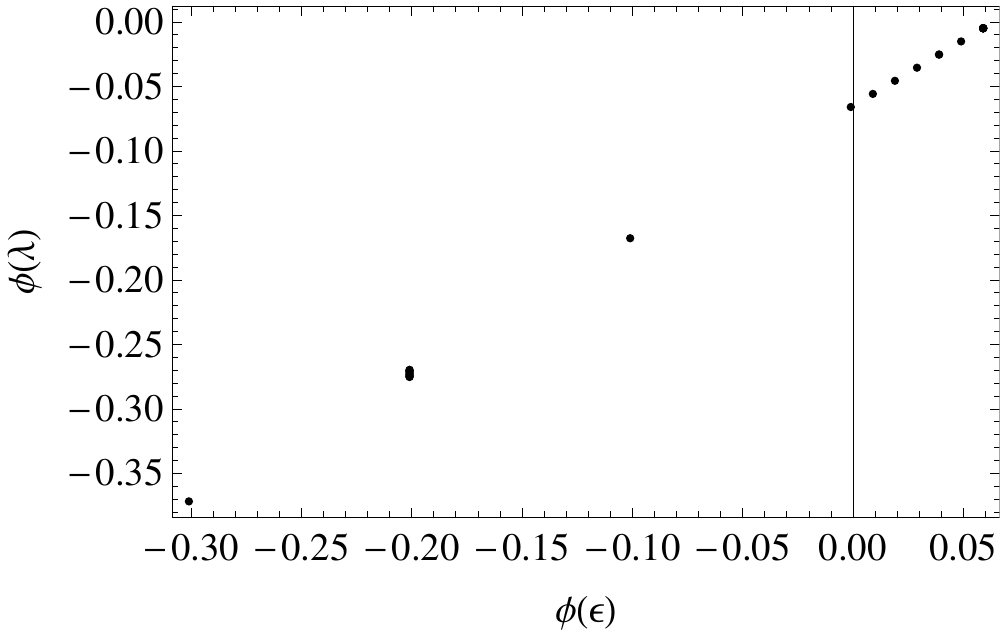}
  \caption{ $\phi(\lambda)$ vs. $\phi(\varepsilon)$ for the randomly chosen 20 x 20 Hamiltonian matrix.
At $\phi(\lambda)=0$  the correct value $\phi(\varepsilon)=0.064$ is obtained.}
\end{figure}

\vspace{5mm}
This alternating procedure must be done in order to ensure $|\lambda|$ is within a small 
range close to the magnitude $\lambda_{ex}$.

Next, consider the case where the lowest lying eigenvalue is real. The 
iterations, however, are  in the complex plane  and will not always
converge to the proper value of $\lambda$. In order to correct for this, the 
potential, $\hat{V}$ is perturbed slightly by  $\delta i*I$, where $\delta<1$ 
and $I$ is the identity matrix, such that $\hat{V}+\delta i*I = \hat{V'}$. Replacing $\hat{V}$ with $\hat{V'}$ 
in the Hamiltonian will result in a ground state with a complex eigenvalue and 
the algorithm as described above can be applied.  FIGS. 4 and 5 show the 
convergence for this case.

\vspace{3mm}
\begin{figure}[htbp]
  \centering
  \includegraphics[scale=1]{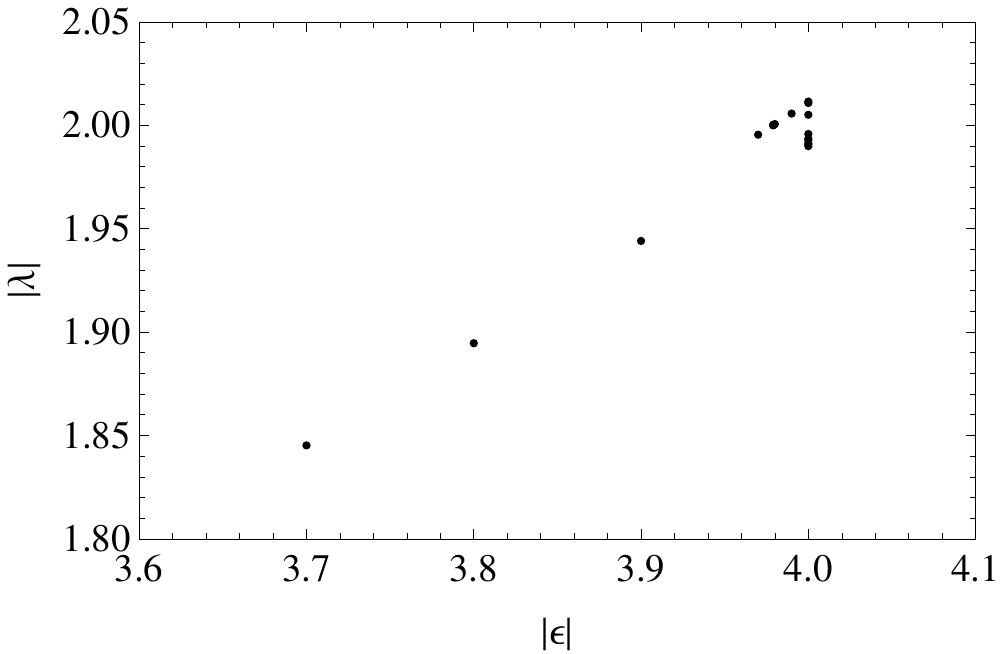}
  \caption{$|\lambda|$ vs. $|\varepsilon|$ for the 20x20 perturbed Hamilton matrix in which
the original (unperturbed) ground state eigenvalue is real. Here
$|\lambda|=2=\lambda_{ex}$ with the corresponding value $|\varepsilon|=3.978$}
\end{figure}

\begin{figure}[h!tbp]
\vspace{-1mm}
  \centering
   \includegraphics[scale=1]{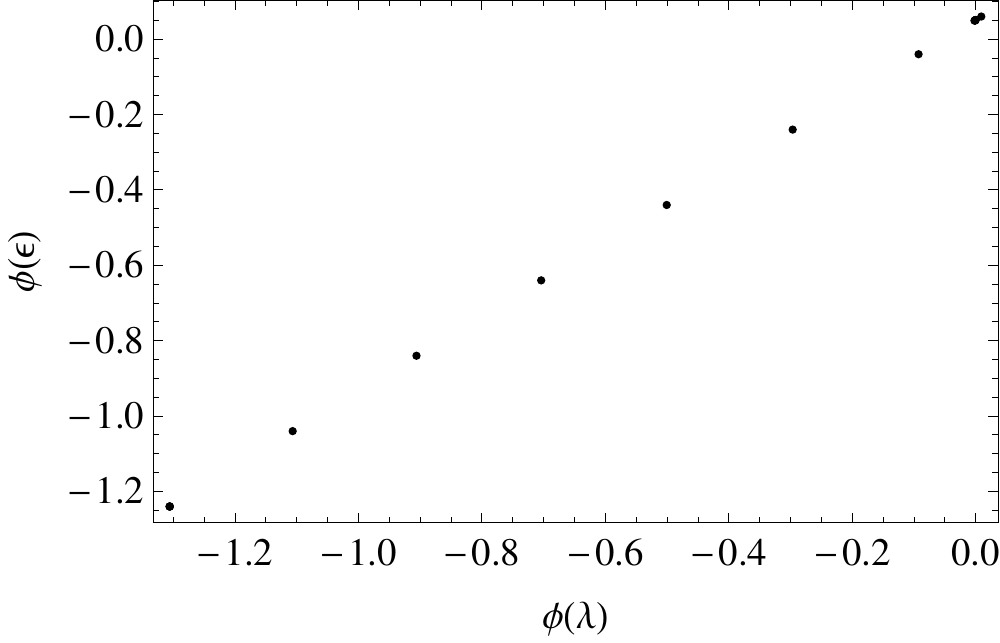}
   \caption{$\phi(\lambda)$ vs. $\phi(\varepsilon)$ for the shifted perturbed 20x20
Hamilton matrix. When $\phi(\lambda)=0$, $\phi(\varepsilon)=0.05$.}
\end{figure}

In the present work we have extended Waxman's algorithm to include 
non-Hermitian Hamiltonian
operators. A convergent iterative scheme is presented to find the lowest lying 
eigenstate of such operators.
For Hamiltonians  whose ground state eigenvalues are real a simple prescription 
is given to guarantee convergence. Excited states
may be obtained from a new start vector in which  the the lower lying eigenstates are projected out of 
the original start vector.


\end{document}